# An investigation into the nano-/micro-architecture of electrospun poly (ε-caprolactone) and self-assembling peptide fibers

Robabeh Gharaei,[1] Giuseppe Tronci,[1, 2] Robert P. Davies,[2] Parikshit Goswami,[1] and Stephen J. Russell[1]

[1] Nonwovens Research Group, School of Design, University of Leeds, Leeds LS2 9JT, United Kingdom
[2] Biomaterials and Tissue Engineering Research Group, School of Dentistry, St. James's University Hospital, University of Leeds, Leeds LS9 7TF, United Kingdom

## ABSTRACT

Self-assembling peptides (SAPs) have the ability to spontaneously assemble into ordered nanostructures enabling the manufacture of 'designer' nanomaterials. The reversible molecular association of SAPs has been shown to offer great promise in therapeutics via for example, the design of biomimetic assemblies for hard tissue regeneration. This could be further exploited for novel nano/micro diagnostic tools. However, self-assembled peptide gels are often associated with inherent weak and transient mechanical properties. Their incorporation into polymeric matrices has been considered as a potential strategy to enhance their mechanical stability. This study focuses on the incorporation of an 11-residue peptide, $P_{11}$-8 (peptide sequence: $CH_3CO$-Gln-Gln-Arg-Phe-Orn-Trp-Orn-Phe-Glu-Gln-Gln-$NH_2$) within a fibrous scaffold of poly (ε-caprolactone) (PCL). In this study an electrospinning technique was used to fabricate a biomimetic porous scaffold out of a solution of $P_{11}$-8 and PCL which resulted in a biphasic structure composed of submicron fibers (diameter of 100-700 nm) and nanofibers (diameter of 10-100 nm). The internal morphology of the fabric and its micro-structure can be easily controlled by changing the peptide concentration. The secondary conformation of $P_{11}$-8 was investigated in the as-spun fibers by ATR-FTIR spectroscopy and it is shown that peptide self-assembly into β-sheet tapes has taken place during fiber formation and the deposition of the fibrous web.

## INTRODUCTION

The effectiveness and elegance of molecular self-assembly in biology has inspired researchers to focus on self-assembling peptide materials as novel nanostructured biomaterials for use in therapeutics. $P_{11}$-8 is an 11 amino acid amphiphilic peptide (with +2 net charge) and in its monomeric natural form is designed to adopt a β-strand conformation. $P_{11}$-8 β-strands can subsequently undergo triggered self-assembly (depending on peptide concentrations, and environmental factors, e.g. pH and temperature) into long β-sheet tapes (single-molecule thick), ribbons (two stacked tapes), fibrils (multiple stacks of ribbons) and fibers (entwined fibrils), without the presence of other conformations such as turns, loops or α-helices. [1, 2]. It also self-assembles and forms self-supporting gels in physiological conditions above a critical concentration [1]. $P_{11}$-8 hydrogel has been studied as a biomaterial because of its potential to mimic the properties of extracellular matrix (ECM) in native tissues. It has shown to be non-cytotoxic to human and murine cells [1, 3]. $P_{11}$-8 also has potential benefits in bone tissue regeneration *in vivo*. However, these protein-based hydrogel biomaterials are intrinsically linked to weak mechanical properties making their handling during implantation challenging and limiting their applicability in large load-bearing tissue defects. Combining the unique properties of self-assembling peptides (SAPs) and synthetic polymers, e.g. via surface modification and coating of poly(ε-caprolactone) (PCL) films or nanofibers with self-assembling peptides [4-6], or incorporation of peptides within PCL fibers during electrospinning [7], is a potential route that can enhance the mechanical properties of SAPs. In the present work, PCL was chosen as a base polymer due to its suitability for various tissue engineering applications including bone regeneration [8] and electrospinning was used as fabrication technique due to several advantages of electrospun fibers such as high surface area to volume ratio and high porosity suitable for tissue engineered scaffolds [9]. It was hypothesized that $P_{11}$-8 can be added into PCL solution prior to electrospinning when it is in monomeric random coil conformation. During electrospinning and solvent evaporation, the peptides may be expected to self-assemble in this environment due to a sharp increase in peptide concentration. The aim of this study was to combine the self-assembled $P_{11}$-8 and PCL in the form of electrospun fibers and make a structurally reinforced, porous and biofunctional scaffold of PCL/$P_{11}$-8 peptide.

**EXPERIMENTAL DETAILS**

$P_{11}$-8 (peptide content ~ 75%, HPLC purity of 96%) was purchased from CS Bio Co. (USA); all other chemicals were purchased from Sigma-Aldrich. The electrospinning solutions were prepared by dissolving 6% (w/w) PCL in hexafluoro-2-propanol (HFIP). Selected amounts (10, 20 or 40 mg) of $P_{11}$-8 were added into 1 mL of the PCL solution. The standard single spinneret electrospinning setup was applied using a 5 mL glass syringe and a 22 gauge blunt tipped needle. The flow rate was 1 mL $h^{-1}$. The distance between the needle and the collector was 18 cm and a 20 kV voltage was used to form smooth and uniform defect-free fibers. The fibers were collected on aluminum foils and were dried at room temperature under vacuum for 7 days. Field emission gun scanning electron microscope (FEGSEM) images of the electrospun samples were taken by sputter coating the samples with platinum (8 nm thickness) and then imaging the samples in a LEO1530 Gemini machine with an accelerating voltage of up to 5 kV and working distance of 3.0-3.5 mm. A thin layer of the fibers was directly electrospun into the transmission electron microscopy (TEM) grids (mounted on aluminum foil) and TEM analysis was performed using a field emission gun transmission electron microscope (FEGTEM) of model type FEI Tecnai TF20. Attenuated total reflection Fourier transform infrared (ATR-FTIR) spectroscopy was carried out on the electrospun fibers using a Perkin Elmer FTIR system and 64 scans were averaged for each spectrum. FTIR of the solutions was carried out before electrospinning using a Nicolet 6700 FTIR. The samples were held onto $CaF_2$ windows and 32 scans were averaged for each spectrum. The HFIP solvent spectrum (blank) was subtracted from the spinning solution spectra.

**DISCUSSION**

**Microscopic morphology of the fibrous scaffolds**

Electrospun $P_{11}$-8/PCL fibers were found to exhibit distinctive fiber morphology when compared to the PCL control sample (Figure 1). The PCL fibers containing $P_{11}$-8 peptide combined two different fiber networks wherein some fibers were of submicron dimensions (diameter of 100-700 nm) and others were nanofibers (diameter of 10-100 nm). A net-like network of small diameter fibrils could also be observed in the higher magnification TEM images (Figure 2), which may be indicative of aggregation of single β-sheet-ordered fibrils within the scaffold. The nanofibers appeared to be either generated from the submicron parent fibers (shown in the TEM image in Figure 2 B) or independently distributed throughout the fibrous structure. The dimensions of the nanofibers are comparable with the diameter of $P_{11}$-8 fibrils in the gel form (10-20 nm) reported in previous studies [1]. Increasing the concentration of $P_{11}$-8 (10, 20 and 40 mg mL$^{-1}$) resulted in more pronounced networks of nanofibrils (shown in the SEM images in Figure 1). As the peptide concentration was found to control the nanoscale architecture of the scaffold, bespoke fiber/fibril morphology could be accomplished to suit the requirements of specific tissue engineering applications.

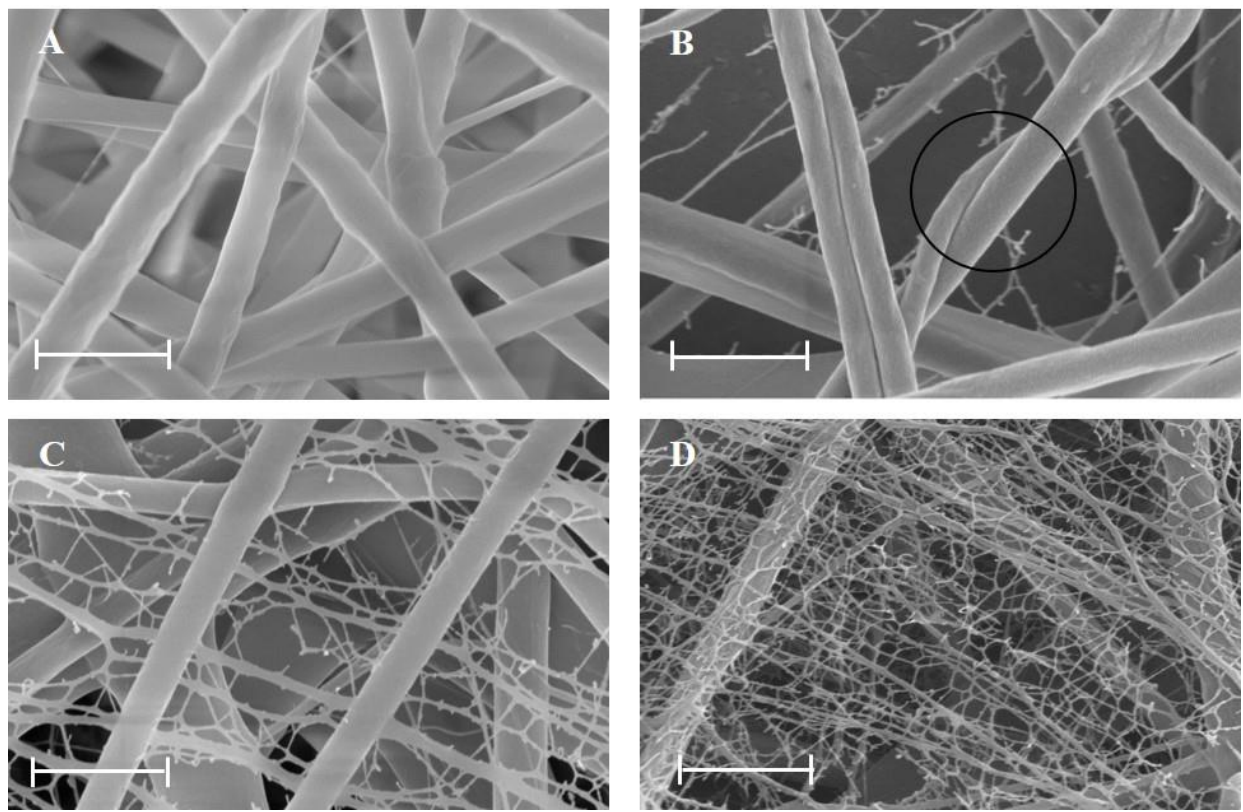

**Figure 1.** Scanning Electron Micrographs of PCL fibers (A), PCL/$P_{11}$-8 with peptide concentration; 10 mg mL$^{-1}$ (B), 20 mg mL$^{-1}$ (C), 40 mg mL$^{-1}$ (D). The scale bar in all the images is 1 micron.

Moreover, single or double twisted fibers were observed in PCL/$P_{11}$-8 samples at the submicron scale (e.g. in Figure 2.A and Figure 1.B respectively) which resembles the molecular orientation of single β-sheet tapes stacking face to face to form twisted fibrils. This may further support the β-sheet conformation of $P_{11}$-8 within the submicron fibers (diameter more than 100 nm) as well. However, this kind of bimodal fiber morphology resulting from the production of a net-like network of nanofibers among the larger submicron fibers has also been linked to the conductivity of the spinning solution [10, 11]. In other studies, presence of a secondary electric field between the molecules in spinning solutions, which can be altered by e.g. addition of salts [12, 13] has been proposed as a synthesizing hypothesis for production of nanofibril networks. In the present study, $P_{11}$-8 peptide has a +2 overall net charge, which can be expected to increase the conductivity of the peptide electrospinning solution compared to a solution of PCL alone, and thereby to provide a plausible hypothetical mechanism for the formation of nanofibrils. Likewise it can be said that as $P_{11}$-8 has amino acids with negatively or positively charged side-chains within its molecular structure, a contributory mechanism for the formation of nanofibrils networks can be the secondary attraction between these charged groups.

The fibers containing peptides show a wider submicron fiber diameter distribution (100-700 nm) compared to the PCL fibers (300-500 nm). This larger fiber diameter distribution in the biphasic networks (10-700 nm) may be beneficial in biomaterials where the staggered release of multiple drugs is required. The biphasic architecture may also allow for a controlled degradation profile of tissue engineering scaffolds since the nanofibers can be expected to degrade first due to their higher surface area to weight ratio. Moreover the nanofibrillar network within the pores may promote homing of endogenous cells within the core of the electrospun scaffold, which is crucial for functional tissue regeneration [14], as well as enhance its mechanical strength [15].

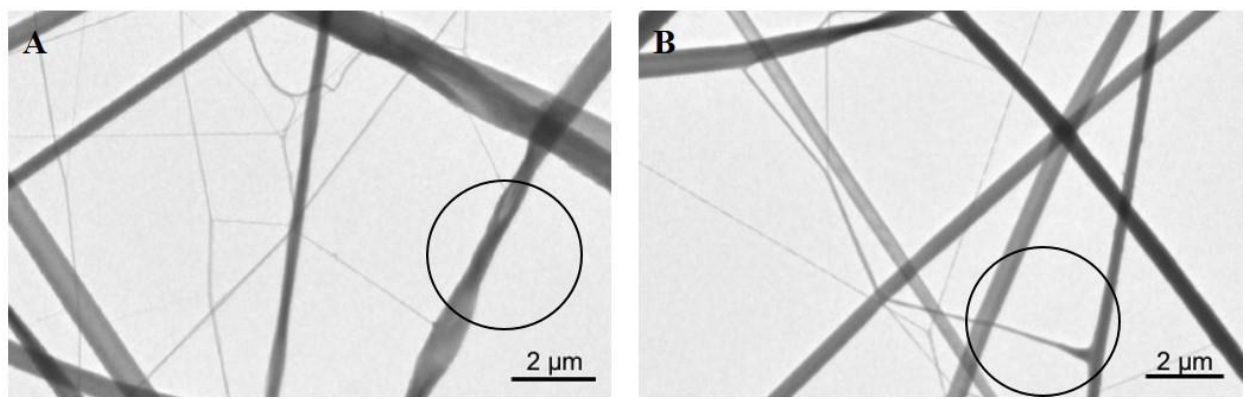

**Figure 2.** TEM images of PCL/$P_{11}$-8 (20 mg mL$^{-1}$ peptide concentration) showing a network of large diameter twisted submicron fibers (A) combined with a nanofibrillar network of small fibers that appear to cross or emerge from the boundaries of the submicron fibers**.**

## Molecular organization of $P_{11}$-8 in the electrospinning solution and in the as-spun fibers

In order to ascertain conformation of peptide in the electrospinning solutions with 20 mg mL$^{-1}$ peptide concentration and then to assess the conformation of the peptide in the resulting as-spun fibers, the materials were characterized using FTIR and ATR-FTIR respectively. The FTIR spectra of PCL/$P_{11}$-8 electrospinning solution (Figure 3.A) show a broad peak at 1650 cm$^{-1}$ confirming a random coil peptide configuration [16]. The ATR-FTIR spectra for electrospun PCL/$P_{11}$-8 fibers are shown in Figure 3.B Besides the strong peak at 1727 cm$^{-1}$ representing the carbonyl stretching of PCL, peaks at 1630 and 1682 cm$^{-1}$ as well as the missing peak at 1650 cm$^{-1}$ give supporting evidence of the predominant antiparallel β-sheet conformation of peptide in the fibers [16]. These FTIR spectra in the amide I′ region of $P_{11}$-8 fibers are comparable with that of strong self-supporting β-sheet hydrogels published in a previous study [17]. The characteristic strong band resulting from carbonyl stretching of PCL and also the band at 1650 cm$^{-1}$ corresponding to peptide in a random coil state are slightly shifted in the PCL/$P_{11}$-8 solution, which is expected due to peptide/PCL/HFIP secondary interactions. Based on these FTIR spectra and previous SEM and TEM observations, it is therefore suggested that peptide assembling into β-sheet is accomplished following electrospinning, ultimately leading to structures of higher hierarchical assembly, i.e. fibrils and fibers (entwined fibrils). Nanofibrils could also be formed soon after the formation of the submicron fibers as the process of solvent evaporation from the fibers and increase in peptide concentration continues in the vicinity of the collector.

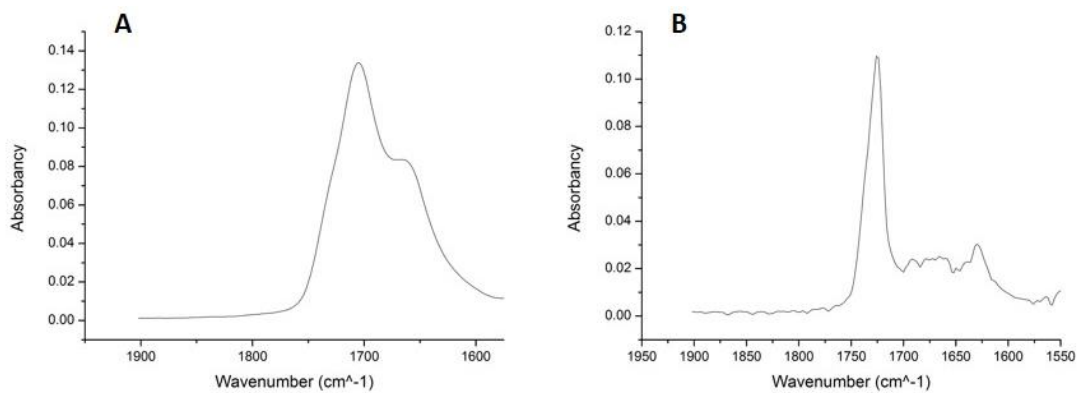

**Figure 3.** Secondary structure analysis FTIR of solution of PCL/$P_{11}$-8 (20 mg mL$^{-1}$ peptide concentration) before electrospinning (A) and ATR-FTIR of resulted as-spun fibers after electrospinning (B).

## CONCLUSIONS

The creation of a biphasic fibrous structure was successfully demonstrated by a one-step electrospinning process with monomeric $P_{11}$-8 peptide and PCL in HFIP. Two mechanisms of fiber formation appeared to contribute to the final architecture, one of which is associated with electrospinning from the submicron up to nanoscale, and the other with peptide self-assembling from the molecular up to the nanoscale. Based on the diameter, morphology and molecular organization of the fibers, it can be proposed that $P_{11}$-8 is self-assembled into self-supporting nanofibrils or into ordered structures within the submicron electrospun fibers. The internal structure of the fabric at nanoscale can be systematically customized by altering the peptide concentration in the electrospinning solution. The peptide incorporated into the fibers displayed β-sheet conformation in contrast to the peptide's monomeric conformation in the electrospinning solution, which confirms that the electrospinning process does not interfere with the molecular self-assembling mechanism of $P_{11}$-8 peptides. The self-assembly of $P_{11}$-8 during electrospinning could be attributed to solvent evaporation and the resultant increase in the peptide concentration. By exploring the biphasic structure at higher magnification, it appeared that some of the secondary nano-networks originated from the parent electrospun fibers as a result of the self-assembly mechanism. Indeed, discontinuities in the fibril-based nano-network would be expected if fibrils were formed during the electrospinning process alone as a result of the high velocity, non-laminar spinning jet streams. Consequently, an *in-situ* mechanism of self-assembling of the peptide is possible following electrospinning due to the formation of intra-molecular hydrogen bonds within the peptide molecule as the solvent evaporates.

## ACKNOWLEDGMENTS

The authors gratefully acknowledge financial support from the Alumni of the University of Leeds for the research scholarship awarded to RG.